\documentclass[journal]{IEEEtran}
\usepackage[T1]{fontenc}
\usepackage{mathrsfs}
\usepackage{bm}
\usepackage{amsmath}
\usepackage{amsthm}
\usepackage{amssymb}
\usepackage{amsfonts}
\usepackage{float}
\usepackage{bbm}
\usepackage{hyperref}
\usepackage[caption=false,font=footnotesize]{subfig}

\usepackage{graphicx}
\usepackage{verbatim}
 \PassOptionsToPackage{normalem}{ulem}
\usepackage{scalerel}

\usepackage{cite}
\usepackage{graphicx,psfrag}
\usepackage[table]{xcolor}
\usepackage{epstopdf}

\usepackage[ruled,vlined]{algorithm2e}

\newtheorem{remark}{Remark}

\begin{document}
\title{Inverse Quickest Change Detection}% for counter-adversarial autonomous systems}

\author{Saqib Abbas Baba, Kumar Vijay Mishra, Lance Kaplan, and Arpan Chattopadhyay

\thanks{S. A. B. and A. C. are with the Electrical Engineering Department, Indian Institute of Technology (IIT) Delhi, India. A. C. is also associated with the Bharti School of Telecommunication Technology and Management, IIT Delhi. Email: \{saqib.abbas, arpanc\}@ee.iitd.ac.in. K. V. M. and L. K. are with the United States U.S. Army DEVCOM Army Research Laboratory, Adelphi, MD 20783 USA. E-mail: \{kvm@ieee.org, lance.m.kaplan.civ@army.mil\}.}
\thanks{A. C. acknowledges support via  grant no. CRG/2022/003707 from Science and Engineering Research Board (SERB), India, project no. RP04860N from SYSTRA MVA Consulting India Private Ltd.
India, grant no. IFC/7150/2023 from Indo-French Centre for the Promotion of Advanced  Research,   and Project No. FT/2024/11/37 from QUALCOMM TECHNOLOGIES, INC, USA.   K. V. M. acknowledges partial support from the National Academies of Sciences, Engineering, and Medicine via the Army Research Laboratory Harry Diamond Distinguished Fellowship and ARL AI and Autonomy for Multi-Agent Systems (ArtIAMAS) program. The views and conclusions contained in this document are those of the authors and should not be interpreted as representing the official policies, either expressed or implied, of the Army Research Laboratory or the U.S. Government. The U.S. and Indian  Governments are authorized to reproduce and distribute reprints for Government purposes notwithstanding any copyright notation herein.}
}

\maketitle

\begin{abstract}
In recent years, \textit{inverse cognition} technique that models intelligent targets capable of evading, estimating, and entrapping cognitive radar systems has garnered growing attention. Prior work has primarily focused on designing inverse stochastic filters that enable a target to infer the tracking accuracy of an interrogating radar. In this work, we shift the focus to the detection aspect of the problem. Specifically, we consider a cognitive radar employing Bayesian \textit{quickest change detection} (QCD) to identify, with minimal delay, a change in the statistical distribution of noisy target observations, subject to false alarm constraints. The radar declares a change once the posterior probability of the target’s entry into its coverage surpasses a predefined threshold. In response, we model an intelligent target as an inverse agent aiming to estimate the radar’s detection time. The target leverages its knowledge of its own state, exact entry time in the radar's coverage, and partial observations of the radar's actions. To this end, we develop an \textit{inverse QCD} framework that uses a particle filter (PF) to approximate the radar’s posterior distribution and compute a detection statistic to infer whether a change detection decision has occurred. We also address the case where the radar’s detection threshold is unknown by introducing an online PF-based estimator, enabling the target to adaptively learn the radar's decision boundary. Furthermore, we compare alternative detection statistics, including the posterior mean of the radar's QCD statistic and a direct likelihood ratio test, analyzing trade-offs between detection delay and false alarm rate. Numerical simulations validate our proposed techniques and demonstrate their robustness under diverse operational scenarios.
\end{abstract}

\begin{IEEEkeywords}
Bayesian quickest change detection, cognitive radar, electronic warfare, inverse cognition, Shriyaev's test.
\end{IEEEkeywords}

\section{Introduction}\label{section:introduction}

The study of detection and inference in adversarial scenarios has garnered significant attention due to its critical applications in domains such as radar surveillance, cognitive electronic warfare, and autonomous systems. A fundamental problem in this context is the quickest change detection (QCD) problem, where a decision-maker seeks to detect a sudden change in the statistical distribution of sequential observations as quickly as possible, while controlling false alarms. A radar system is a canonical example, where the appearance and disappearance of a target can be  modeled as a change in the distribution of received signals at the radar. However, the traditional QCD framework predominantly focuses on the radar's perspective, optimizing its detection performance. This paper shifts the focus to the target, positioning it as an “inverse agent” seeking to infer the radar's detection strategy and stopping time, i.e., the time the radar detects the presence of the target and starts taking counter-measures against the target. This approach, termed \textbf{inverse QCD}, has profound applications towards the development of stealthy, adaptive, and cognitive target behavior in radar-evading scenarios.  This problem becomes challenging if the target’s observations of radar actions are noisy or indirect, and if the radar’s detection strategy  is partially or entirely unknown. To address these challenges,  we propose a novel particle filtering and learning based inverse QCD algorithm named IQCD, which performs  inverse QCD in the Bayesian setting by suitably updating a posterior probability of change,  and also handles both known and unknown thresholds used by the radar. Numerical results demonstrate that IQCD offers a tractable and robust solution to the inverse QCD problem. 

\subsection*{Prior Art}

The theory of QCD (\cite{poor2008quickest, tartakovsky2014sequential}) has been extensively studied in the context of sequential analysis and statistical signal processing. Shiryaev \cite{shiryaev1963optimum} first formalized the Bayesian QCD framework, where the decision maker  starts with a prior belief on the distribution of the change instant, and has to decide when to stop gathering observation and declare a change. The goal of the decision maker is to minimize the expected detection delay subject to a constraint on false alarm probability. In the test proposed by Shiryaev, the decision maker   maintains and keeps updating a posterior belief on the change having already occurred, and declares a change when this belief crosses a suitable threshold. On the other hand, Lorden \cite{lorden1971procedures} and Pollak \cite{pollak1985optimal} introduced non-Bayesian formulations under adversarial change instant and pre-change observations, where the goal was to optimize {\em worst-case}  detection delay under false alarm rate constraints. The standard  algorithm for non-Bayesian QCD is the Cumulative Sum (CUSUM) test \cite{page1954continuous}, which is essentially a   likelihood ratio test.  

This paper primarily focuses on inverse QCD problems in the larger context of inverse cognition for autonomous systems. Inverse cognition is typically performed by an {\em inverse agent} which observes the (possibly adaptive) actions of a forward agent and infers the forward agent's knowledge about the inverse agent. One obvious example scenario is the interaction between a cognitive radar and an intelligent target. While the cognitive radar sequentially collects observations about the target and adaptively optimizes its detection strategy, an intelligent target equipped with suitable sensors can observe a noisy version of the radar's action and infer the radar's belief about the target's state and as well as the parameters and strategies (e.g.,  detection threshold, sensing strategy, or transmitted waveform) used by the radar. However, the concept of {\em inverse problems} has been widely studied in physics, machine learning, and control systems\cite{tarantola2005inverse}. 

On the other hand, inverse stochastic  filtering, another form of inverse cognition, has been extensively  studied for nonlinear sensing systems \cite{singh2023inverse1, singh2023inverse2, singh2024inversecubature, singh2024inverseunscented, singh2025inverseparticle}. In this framework, a cognitive radar tracks a target's state sequentially by using stochastic filters such as extended Kalman filter, unscented Kalman filter, particle filter etc., and simultaneously keeps adapting its action based on the inferred state of the target. The target sequentially observes noisy versions of the radar's actions, and uses these observations to perform inverse filtering to infer the radar's estimate of the target's state.   

Meanwhile, Pattanayak et al.\cite{pattanayak2023necessary} investigate inverse reinforcement learning in Bayesian stopping time problems by establishing conditions under which an agent’s observed actions can be deemed cost-optimal, thereby enabling the identification of set-valued cost functions. They also propose meta-cognitive radar techniques\cite{pattanayak2023meta} designed to obscure the radar’s true utility function from an adversary seeking to reverse-engineer its cognition, thus illustrating how inverse learning insights can be leveraged to maintain strategic advantage in electronic warfare settings. 

Additionally, authors in \cite{krishnamurthy2021adversarial} consider three closely related adversarial inference problems in cognitive radars, including inverse tracking (to estimate the radar’s own estimate of the target), identifying cognition via revealed preference tests, and designing smart interference strategies. By treating the radar as a constrained utility maximizer, this framework provides a powerful microeconomics-based perspective on how the target can deduce the radar's internal objectives and sensing accuracy.

Another notable direction is inverse sequential hypothesis testing (SHT) addressed in \cite{pattanayak2020inverse}, where an inverse agent observes a Bayesian decision-maker’s sequential detection actions and uses these observations to infer its detection costs. The authors use concepts from behavioral economics and Bayesian revealed preference theory to show that a Bayesian agent performing SHT can be viewed as a rationally inattentive utility maximizer. This line of work is closely related to inverse quickest change detection; however, its objective is to infer the underlying cost structure governing the stopping rule, rather than the stopping time itself.

The notion of cognitive targets, which adaptively learn and respond to radar behavior, has been introduced in cognitive electronic warfare. These targets typically rely on predefined models of the radar's detection process to guide their actions, such as minimizing exposure time or maximizing stealth. For example, a target might analyze the radar's waveform patterns to predict detection zones and adjust its flight path accordingly, thereby minimizing exposure time. In contrast, the proposed inverse QCD framework allows the target to explicitly infer the radar’s stopping time by leveraging partial observations of its actions. This direct estimation of the radar’s stopping time and detection strategy enables more precise and dynamic target responses in adversarial scenarios.

Consider a cognitive target operating in an adversarial environment. Prior to being detected, the target may choose to remain silent, preserve energy, or maintain its nominal trajectory. Once the radar detects the target, the target may need to rapidly initiate evasive maneuvers, activate electronic countermeasures, deploy decoys, or leave the surveillance region. In such scenarios, delayed inference of the radar's detection decision may significantly reduce the effectiveness of the target's response, whereas premature inference may lead to unnecessary expenditure of resources or compromise the target's mission objectives. Consequently, the target benefits from inferring the radar's detection decision as quickly as possible while maintaining a controlled false alarm probability. This naturally leads to an inverse quickest change detection problem, in which the target seeks to infer the radar's stopping time subject to its own false alarm constraints.

Despite extensive research on QCD and its applications in radar systems, the inverse QCD problem has received limited to no attention. Existing works on cognitive radar and cognitive targets primarily focus on optimizing radar detection or target evasion strategies without delving into the inference of radar's detection activities at the target's end. This paper addresses this gap by:
\begin{enumerate}
\item Proposing  {\em non-standard} definitions   of false alarm probability and conditional mean detection delay for inverse QCD, and using them as detection performance metrics at the target.
\item Formulating the inverse QCD problem, where the target seeks to estimate the radar's stopping time under both known and unknown radar threshold scenarios.
\item Applying a particle filter-based framework for the target to compute certain {\em non-standard} posterior probabilities and the radar's  detection  threshold, in order to facilitate inference of the radar's stopping time by the target.
\item Extending the proposed framework to account for uncertainty in the radar observation model and the action-observation model (at the target) through a multiple-model IQCD (MM-IQCD) formulation.
\end{enumerate}
The performance of the proposed IQCD algorithm is then demonstrated to be superior via extensive numerical comparison against competing algorithms. 

In the remainder of this paper, we formalize this inverse QCD framework and detail a particle filtering–based estimation scheme. After formulating the radar model in Section~\ref{section:System_Model}, we discuss the target's perspective for inverse QCD in Section~\ref{section:target's IQCD}. Section~\ref{section:PF} presents our particle filtering approach for real-time estimation of the radar’s stopping time, including an extension for threshold inference and model uncertainty. Numerical results in Section~\ref{section:numerical_results} demonstrate that the proposed method accurately predicts stopping times under diverse noise conditions. Finally, we provide concluding remarks in Section~\ref{section:conclusion}.

The proposed framework is validated through numerical experiments, demonstrating accurate inference of the radar's stopping time, robustness to threshold and model uncertainty, and improved performance over competing baseline methods.

\section{System Model: Forward Quickest Change Detection Procedure for the Radar}\label{section:System_Model}

\subsection{Radar's Detection of the Target}
In this section, we describe the radar’s forward quickest change detection (QCD) procedure for monitoring the state of a potentially incoming target. The radar makes sequential noisy observations \( Y_t \) about the target's state at discrete time steps \( t = 1, 2, 3, \ldots \). At some unknown time instant \( \tau \), the target enters the sensing region of the radar, causing a change in the statistical properties of the observations. Specifically, the observation \( Y_t \) follows a pre-change distribution \( f_0 \) and a post-change distribution \( f_1 \):
\begin{equation*}
    Y_t \sim 
    \begin{cases}
        f_0(\cdot), & t<\tau,\\
        f_1(\cdot), & t \geq \tau.
    \end{cases}
\end{equation*}
The change point \( \tau \) is unknown to the radar and is modeled as a random variable with a geometric prior distribution which is common in Bayesian QCD literature\cite{poor2008quickest}. The probability that the change occurs at time \( k \) is given by
\begin{equation}\label{eq:change_point_prior}
    \mathbb{P}(\tau = k) = (1 - p)^{k - 1} p, \quad \text{for } k = 1, 2, 3, \ldots,
\end{equation}
where \( p \in (0, 1) \) is the known parameter of the geometric distribution, representing the conditional probability that the change occurs at a given time step, given that no change has occurred up to that point. The target, however, has complete knowledge of the change point \( \tau \).

The radar's objective is to detect the change as quickly as possible after it occurs, while controlling the probability of a false alarm (i.e., declaring a change before it happens). Formally, the radar seeks to solve the following optimization problem\cite{shiryaev1963optimum}:
\begin{equation}\label{eq:radar_optimization_problem}
    \begin{aligned}
        & \underset{T}{\text{minimize}} && \mathbb{E}\left[ (T - \tau)^+ \right] \\
        & \text{subject to} && \mathbb{P}(T < \tau) \leq \alpha,
    \end{aligned}
\end{equation}
where the objective function is essentially the mean detection delay, \( T \) is a stopping time representing the time at which the radar declares that the change has occurred, \( x^+ = \max\{x, 0\} \), and \( \alpha \) is the specified probability of false alarm. The operator $\mathbb{E}[\cdot]$ denotes expectation, and $\mathbb{P}(T<\tau)$ is the probability of a false alarm.

A well-known solution to \eqref{eq:radar_optimization_problem} in the Bayesian setting was proposed by Shiryaev (see \cite{shiryaev1963optimum}). It maintains a posterior probability $\pi_t$ that the change has already occurred by time $t$, given the observation sequence $Y_{1:t} = (Y_1, Y_2, \ldots, Y_t)$. 
\begin{equation}\label{eq:posterior_probability}
    \pi_t = \mathbb{P}(\tau \leq t \mid Y_{1:t}).
\end{equation}

The update of the posterior probability is performed using Bayesian recursion. The radar updates \( \pi_t \) as follows:
\begin{equation}\label{eq:posterior_update}
    \pi_{t} = \frac{\tilde{\pi}_{t - 1} L(Y_t)}{\tilde{\pi}_{t - 1} L(Y_t) + (1 - \tilde{\pi}_{t - 1})} , 
\end{equation}
where
\begin{equation}\label{eq:prior_update}
    \tilde{\pi}_{t - 1} = \pi_{t - 1} + (1 - \pi_{t - 1}) p,
\end{equation}
and
\begin{equation}\label{eq:likelihood_ratio}
    L(Y_t) = \frac{f_1(Y_t)}{f_0(Y_t)}
\end{equation}
is the likelihood ratio of the observation \( Y_t \).

Once the radar has computed $\pi_t$, it declares detection at the first time $t$ such that $\pi_t$ exceeds a threshold $h_{\text{rad}}$. That is, the stopping time \( T \) is defined as
\begin{equation}\label{eq:radar_stopping_time}
    T = \inf \left\{ t \geq 1 : \pi_t \geq h_{\text{rad}} \right\}.
\end{equation}
The threshold \( h_{\text{rad}} \) is chosen to satisfy the false alarm constraint \( \mathbb{P}(T < \tau) \leq \alpha \) with equality.

\subsection{Radar's Action}
At each time $t$, after updating its posterior probability $\pi_t$, the radar takes an action \(g_t\) that reflects its operational decision, such as altering transmission power, beam direction, waveform parameters, or initiating target-tracking protocols. In general, we model the radar's action as a random variable whose distribution depends on the posterior probability \(\pi_t\):
\begin{equation}\label{eq:radar_action}
    g_t \sim p(g_t \mid \pi_t),
\end{equation}
where \(g_t\in\mathcal{G}\), with \(\mathcal{G}\) denoting the radar's action space. This formulation allows the radar's action to depend on its belief state in an arbitrary manner.

In this work, however, we consider the practically relevant case in which the radar changes its operating mode once its posterior exceeds the detection threshold ($h_{\text{rad}}$). Consequently, the target observes two distinct action-observation models corresponding to the pre- and post-stopping regimes, leading to the observation model to be introduced in Section~\ref{subsection:Problem-setup}.

This forward QCD model of the radar’s decision-making process forms the basis for the inverse QCD framework described in the next sections. By observing the radar’s actions and leveraging partial knowledge of its QCD mechanism, the target seeks to infer the radar’s stopping time $T$ and, in some scenarios, the unknown threshold $h_{\text{rad}}$.

\section{Target's Inverse QCD Problem}\label{section:target's IQCD}

Having described the radar’s quickest change detection (QCD) procedure, we now shift the perspective to the target, which aims to infer when the radar first declares detection. Unlike the radar which observes $\{Y_t\}_{t\geq1}$, the target indirectly observes the radar’s action $\{g_t\}_{t\geq 1}$. 

\subsection{Problem setup}\label{subsection:Problem-setup}
The target is aware that the radar's actions depend on its detection strategy and can observe these actions (possibly with some noise). Let $\tilde{g}_t$ denote the target's noisy observation of the radar's action $g_t$. The observation model at the target is characterized by the conditional probability distribution $p(\tilde{g}_t \mid \pi_t)$. In addition, the target knows its own state $X_t \in \{0,1\}$ which indicates whether the target is within the sensing range of the radar. This target state $X_t$ affects how the radar’s posterior $\pi_t$ evolves via the radar’s measurements $Y_t$. The observation model at the target is defined as: 
\begin{equation}\label{eq:observation_model}
p(\tilde{g}_t \mid \pi_t) = 
\begin{cases} 
p_0(\tilde{g}_t), & \text{if } \pi_t < h_{\text{rad}}, \\
p_1(\tilde{g}_t), & \text{if } \pi_t \geq h_{\text{rad}}, 
\end{cases} 
\end{equation} 
where $p_0(\cdot)$ and $p_1(\cdot)$ characterize the likelihood of observed action $\tilde{g}_t$ when $\pi_t$ is respectively below or above $h_{\text{rad}}$. Here $p_0(\cdot)$ and $p_1(\cdot)$ are known to the target (e.g., Gaussians with specified parameters), while $h_{\text{rad}}$ may be known or, if unknown, is estimated as in Section~\ref{section:PF}. Furthermore, we assume that the radar's observation models \( f_0 \) and \( f_1 \) are known to the target as well. This allows the target to accurately simulate the radar's observation process. 

\subsection{Inverse QCD formulation}
We denote by $\hat{T}$ the target’s estimate of $T$. The target seeks to detect the event $\{T \leq t\}$ as early as possible, subject to a constraint on its own false alarm probability. Formally, the target seeks to solve:
\begin{equation}\label{eq:target_optimization_problem}
    \begin{aligned}
        & \underset{\hat{T}}{\text{minimize}} && \mathbb{E}\left[ (\hat{T} - T)^+ \mid T \geq \tau \right] \\
        & \text{subject to} && \mathbb{P}(\hat{T} < T \mid T \geq \tau) \leq \hat{\alpha},
    \end{aligned}
\end{equation}
where $\tau$ is the (known-to-target) time at which the target actually enters the radar’s region, and \( \hat{\alpha} \) is the target's constraint on the probability of false alarm. The conditioning on $\{T \geq \tau\}$ reflects that the radar’s detection time cannot precede the actual change time $\tau$, that is, the radar does not raise a false alarm. Note that in this formulation, a false alarm at the target occurs when it declares that the radar has detected the change before the radar actually does, given that the radar has not falsely declared a change. 

\subsection{Target's posterior probability and stopping rule}

To address the problem in \eqref{eq:target_optimization_problem}, the target maintains the following posterior:
\begin{equation}\label{eq:target_posterior}
    \lambda_t = \mathbb{P}(T \leq t \mid T \geq \tau, X_{1:t}, \tilde{g}_{1:t}).
\end{equation}

By invoking conditional probability relations including the Bayes' theorem, $\lambda_t$ can be written as:
\begin{align}
    \lambda_t &= \frac{\mathbb{P}(\tau \leq T \leq t \mid I_t)}{\mathbb{P}(T \geq \tau \mid I_t)} \nonumber \\
    &= \frac{\mathbb{P}(T \geq \tau \mid I_t) - \mathbb{P}(T > t \mid I_t)}{\mathbb{P}(T \geq \tau \mid I_t)} \nonumber \\
    &= 1 - \frac{\mathbb{P}(T > t \mid I_t)}{\mathbb{P}(T \geq \tau \mid I_t)} \nonumber \\
    &= 1 - \frac{\mathbb{P}(\pi_{t'} < h_{\text{rad}} \text{ for all } t' \leq t \mid I_t)}{\mathbb{P}(\pi_{t'} < h_{\text{rad}} \text{ for all } t' < \tau \mid I_t)}, \label{eq:lambda_expression}
\end{align}
where \( I_t = \{ X_{1:t}, \tilde{g}_{1:t} \} \) denotes the information available to the target at time \( t \), and \( h_{\text{rad}} \) is the radar's detection threshold defined in \eqref{eq:radar_stopping_time}.

The target then applies a threshold rule on $\lambda_t$ to decide when to declare that $T$ has occurred:
\begin{equation}\label{eq:target_stopping_time}
    \hat{T} = \inf \left\{ t \geq 1 : \lambda_t \geq h_{\text{tar}} \right\}.
\end{equation}
where $h_{\text{tar}}$ is chosen to satisfy \( \mathbb{P}(\hat{T} < T \mid T \geq \tau) = \hat{\alpha} \). The PFA at the target depends on $h_{\text{rad}}$ (since both $\tilde{g}_t$ and $T$ depend on it). In practice, we can run Monte Carlo calibration: run simulations to estimate the target’s false-alarm probability $\mathbb{P}(\hat T<T\mid T\ge\tau)$ and tune $h_{\text{tar}}$ until the constraint holds with equality (i.e., equals $\hat{\alpha}$).

Computing the posterior probability \( \lambda_t \) analytically is challenging due to the additional conditioning and the unknown prior on the change point for the target. Therefore, the target employs a particle filter to approximate \( \lambda_t \), as discussed in the next section.

\section{Particle Filter Implementation}\label{section:PF}

Particle filtering, a sequential Monte Carlo (SMC) method, has emerged as a powerful tool for Bayesian inference in nonlinear and non-Gaussian systems\cite{doucet2000sequential}. By approximating a posterior distribution of interest using a set of weighted particles, particle filtering provides a computationally efficient way to approximate complex distributions. Particle filters have shown success in signal processing tasks where closed-form solutions to posterior updates do not exist or are computationally prohibitive. By capturing the posterior distribution over time, particle filtering is well-suited to the inverse QCD problem, wherein the target must track the radar’s evolving detection statistic (and possibly threshold) under uncertainty. 

In the forward QCD problem, the radar updates $\pi_t$ directly via its observations. However, for inverse QCD, the target must simulate or reconstruct the radar’s posterior from partial, noisy observations $\{\tilde{g}_t\}_{t\geq 1}$. Each particle in the PF represents a plausible state of the radar’s belief, along with an associated weight that signifies consistency with the target’s observations and knowledge. Periodic resampling mitigates weight degeneracy, ensuring the particle set remains representative of the underlying distribution.

Since $\lambda_t$ (the target’s posterior probability that the radar has stopped) depends on whether $\pi_{t'}$ stays below $h_{\text{rad}}$ up to $t$, the particle filter effectively tracks $\{\pi_t\}$ over time. Using the weighted particle system at time $t$, the target estimates the two probabilities on the right-hand side of \eqref{eq:lambda_expression} (via the sums in \eqref{eq:prob_approximation}–\eqref{eq:prob_tau_approximation}) and then computes $\lambda_t$ from \eqref{eq:lambda_expression}.

\subsection{Particle Filter Framework}\label{subsection:PF_framework}

\paragraph{Particle representation} 
Let \( \{ (\pi_t^i, w_t^i) \}_{i=1}^N \), where \( N \) is the number of particles, denote the set of particles and their weights at time $t$. Each particle $\pi_t^i \in [0,1]$ represents a possible value of the radar’s current posterior probability that the change has occurred. The weight $w_t^i \geq 0$ is an importance weight proportional to the posterior mass assigned to particle $i$ (see \cite{doucet2011oxford}).

\paragraph{Initialization}
At \( t = 0 \), target assumes the initial radar belief is $\pi_0=0$. Hence, the particles could all be initialized at (or near) zero and the weights are set uniformly as:
        \[\pi_0^i = 0 \quad \text{for } i=1,\ldots,N, \quad \text{and } w_0^i = \frac{1}{N}\]
Alternatively, if the target has an estimate of the parameter $p$, the particles $\{\pi_0^i\}_{i=1}^N$ can be initialized with values drawn uniformly from $[0,\hat{p}]$, where $\hat{p}$ is the target’s nominal estimate of $p$ (e.g., from prior calibration or historical radar logs).

\paragraph{Particle Propagation (Prediction Step)}
At each time \( t \), the target simulates the radar’s Bayesian update to propagate each particle $\pi_{t-1}^i \rightarrow \pi_t^i$. Since the radar’s posterior update (Section~\ref{section:System_Model}) depends on 
$Y_t$ (the radar’s measurement) and the geometric prior parameter $p$, the target must generate a synthetic radar observation $Y_t^i$ for each particle $i$. Specifically,
\begin{equation}\label{eq:synthetic_observation}
    Y_t^i \sim
    \begin{cases}
        f_0(Y_t \mid X_t), & \text{if } t < \tau, \\
        f_1(Y_t \mid X_t), & \text{if } t \geq \tau.
    \end{cases}
\end{equation}
where $X_t$ is the target's known internal state.

The target then applies the usual Shiryaev recursion to each particle (similar to equation \eqref{eq:posterior_update}):
\begin{equation}\label{eq:particle_propagation}
    \pi_t^i = \frac{\tilde{\pi}_{t-1}^i L(Y_t^i)}{\tilde{\pi}_{t-1}^i L(Y_t^i) + (1 - \tilde{\pi}_{t-1}^i)},
\end{equation}
where
\begin{equation}\label{eq:tilde_pi}
    \tilde{\pi}_{t-1}^i = \pi_{t-1}^i + (1 - \pi_{t-1}^i) \hat{p},
\end{equation}
and \( L(Y_t^i) = \frac{f_1(Y_t^i)}{f_0(Y_t^i)} \) is the likelihood ratio.

\paragraph{Weight Update (Correction Step)}
Next, the target corrects each particle’s weight based on the likelihood of the observed action \( \tilde{g}_t \) given \( \pi_t^i \).
\begin{equation}\label{eq:weight_update}
    w_t^i \propto w_{t-1}^i \cdot p(\tilde{g}_t \mid \pi_t^i).
\end{equation}
This is the standard particle filter weight update, i.e., previous weight times the current observation likelihood (see, e.g., \cite{doucet2000sequential, doucet2011oxford}).

The conditional probability \( p(\tilde{g}_t \mid \pi_t^i) \) is computed using the observation model in \eqref{eq:observation_model}. Since the target knows \( \pi_t^i \), it can evaluate \( p(\tilde{g}_t \mid \pi_t^i) \) directly.

Finally, the weights are normalized as follows:
\begin{equation}\label{eq:weight_normalization}
    w_t^i = \frac{w_t^i}{\sum_{j=1}^N w_t^j}, \;\ \forall i.
\end{equation}
This step “corrects” the prior belief about $\pi_t^i$ by emphasizing particles whose predicted radar action is consistent with the target’s actual observation $\tilde{g}_t$.

\paragraph{Resampling}
As time progresses, the particle weights can become degenerate, with only a few particles carrying most of the total weight. To mitigate this, we monitor the effective sample size (ESS), defined as
\[
N_{\text{eff}} = \frac{1}{\sum_{i=1}^N (w_t^i)^2}.
\]
When $N_{\text{eff}}$ falls below a predefined threshold (e.g., $N/2$), resampling is performed:
\[
\{\pi_t^i, w_t^i\}_{i=1}^N \leftarrow \text{Resample}(\{\pi_t^i,  w_t^i\}_{i=1}^N).
\]
After resampling, the weights are reset to $w_t^i = 1/N$ for all $i$.

\subsection{Approximating \texorpdfstring{$\lambda_t$}{lt} and other posterior quantities}

Using the final set of particles at time $t$, the target approximates probabilities relevant to $\lambda_t$ in \eqref{eq:lambda_expression}. We have:
\begin{eqnarray}\label{eq:prob_approximation}
\mathbb{P}(\pi_{t'}<h_{\text{rad}}, \forall t'\leq t |I_t) \approx \sum_{i=1}^{N} w_t^i \mathbbm{1}_{\{\pi_{t'}^i<h_{\text{rad}} \text{ } \forall \text{ } t' \leq t\}}
\end{eqnarray}
and
\begin{eqnarray}\label{eq:prob_tau_approximation}
\mathbb{P}(\pi_{t'}<h_{\text{rad}}, \forall t'\leq \tau |I_t) \approx \sum_{i=1}^{N} w_t^i \mathbbm{1}_{\{\pi_{t'}^i<h_{\text{rad}} \text{ } \forall \text{ } t' \leq \tau\}}
\end{eqnarray}
where \( \mathbbm{1}_{\{\cdot\}}\) is the indicator function. Substituting these approximations into \eqref{eq:lambda_expression}, the target computes the approximation of \( \lambda_t \).

The events in \eqref{eq:prob_approximation} and \eqref{eq:prob_tau_approximation} are path–dependent (they involve the whole trajectory $\pi_{1:t}$). At time $t$, we maintain per-particle indicators (updated recursively), and the current weights $w_t^i$, updated recursively via \eqref{eq:weight_update}, already reflect all data up to $t$. Consequently, the sums \(\sum_i w_t^i \mathbbm{1}_{\{\cdot\}}\) in \eqref{eq:prob_approximation}–\eqref{eq:prob_tau_approximation} are standard SMC estimates of the posterior path probabilities in definition of $\lambda_t$ (see also \cite{doucet2011oxford}). Note that these are approximations because they rely on a finite number of particles.
\begin{remark}[Consistency of particle approximation]
Assume that the particle filter employs the correct state-space model, i.e., the assumed transition and observation models coincide with the true models. Under the standard regularity conditions for sequential Monte Carlo methods (see, e.g., \cite{doucet2000sequential,doucet2011oxford}), the particle approximations in
\eqref{eq:prob_approximation}--\eqref{eq:prob_tau_approximation}
converge almost surely to the corresponding posterior probabilities as
\(N\rightarrow\infty\).
\end{remark}
This result follows directly from the standard consistency theory for particle filters. Under model mismatch, such as \(\hat p\neq p\), increasing the number of particles reduces only the Monte Carlo approximation error, while the discrepancy due to model mismatch remains.

The above particle filter framework enables the target to track the radar’s evolving posterior under uncertainty. By carefully modeling synthetic radar observations, updating particle weights via the target’s own action observations $\{\tilde{g}_t\}$, and resampling, the target can approximate its inverse QCD statistic \( \lambda_t \).

\begin{algorithm}[!t]
\caption{IQCD (inverse QCD, unknown $h_{\mathrm{rad}}$ at the target)}
\label{alg:iqcd_online_joint}
\SetKwInOut{Input}{Input}
\SetKwInOut{Output}{Output}

\Input{Time horizon $K$, number of particles $N$, threshold range $[h_{\min},h_{\max}]$, nominal $\hat p$}
\Output{Target stopping time $\hat T$}

\textbf{Initialize:}
\begin{itemize}
  \item For $i=1,\dots,N$:
  \[
    \pi_0^i \sim U(0,\hat p),\quad
    h_0^{\,i} \sim U(h_{\min},h_{\max})\]
   \[ w_0^i \leftarrow \tfrac{1}{N},\quad
    \texttt{B}_{\tau,0}^i \leftarrow \texttt{True},\quad
    \texttt{B}_{t,0}^i \leftarrow \texttt{True}.
  \]
\end{itemize}
Set $\hat T \leftarrow \varnothing$.

\For{$t=1,\dots,K$}{
    \textbf{Propagate state (per particle):} Update $\pi_t^i$ from $\pi_{t-1}^i$ via the Bayesian recursion using \eqref{eq:particle_propagation}.\\
  \textbf{Weight update} (from \eqref{eq:weight_update}):
  \[
      w_t^i \propto w_{t-1}^i\; p\!\left(\tilde g_t \mid \pi_t^i,\, h^i_{\text{rad},t}\right).
  \]
  \textbf{Normalize weights:}
  \[
      w_t^i \leftarrow \frac{w_t^i}{\sum_{j=1}^N w_t^j}, \qquad \forall i.
  \]
    \textbf{Update Booleans:}
    \begin{itemize}
      \item If $t < \tau$: \(\texttt{B}_\tau^i \leftarrow \texttt{B}_\tau^i \land (\pi_t^i < h^i_{\text{rad},t}) \quad \forall i.\)
      \item \(\texttt{B}_t^i \leftarrow \texttt{B}_t^i \land (\pi_t^i < h^i_{\text{rad},t}) \quad \forall i.\)
    \end{itemize}
    \textbf{Resampling step:} \\
    \textbf{ESS} 
    \[
    N_{\text{eff}} = \frac{1}{\sum_{i=1}^N (w_t^i)^2}
    \]
    \If{$N_{\text{eff}} < N/2$}{
        Resample particles $\{\pi_t^i, h_{\text{rad},t}^i, \texttt{B}_\tau^i, \texttt{B}_t^i\}$ according to weights $w_t^i$.\\
        Reset weights: $w_t^i \leftarrow 1/N,\ \forall i$.
    }
  \textbf{Stopping statistic} (from \eqref{eq:lambda_expression}):
  \[
    \lambda_t \;=\;
    \begin{cases}
      1 - \dfrac{\sum_i w_t^i \texttt{B}_t^i}{\sum_i w_t^i \texttt{B}_\tau^i}, & \text{if } \sum_i w_t^i \texttt{B}_\tau^i > 0,\\[8pt]
      0, & \text{otherwise.}
    \end{cases}
  \]
  \textbf{Decision rule:} \textbf{if } $\lambda_t \ge h_{\mathrm{tar}}$ \textbf{ then } set $\hat T \leftarrow t$ and \textbf{break}.\\
   
  \textbf{Threshold estimate:}
  $\displaystyle\hat h_{\mathrm{rad},t} \;=\; \sum_{i=1}^N w_t^i h_{\mathrm{rad},t}^i.$
}
\textbf{If no trigger:} set $\hat T \leftarrow K\!+\!1$ (no alarm by horizon).\\
\textbf{Return:} $\hat T$
\end{algorithm}

\subsection{Unknown Threshold at the Target}\label{subsection:unknown_threshold}
In practical scenarios, the target may not know the radar’s detection threshold \(h_{\text{rad}}\). Hence, it must estimate \(h_{\text{rad}}\) online while inferring the radar’s stopping time \(T\).
Recall that in the known-threshold setting, each particle \(\pi_t^i\) tracks the radar’s posterior belief at time \(t\).
When \(h_{\text{rad}}\) is also unknown, we augment each particle with a threshold hypothesis, so the particle state is \((\pi_t^i, h^i)\) with weight \(w_t^i\) (see Algorithm~\ref{alg:iqcd_online_joint}).

We assume a prior on the threshold, e.g., $h_0^i \sim U(h_{\min},h_{\max})$ with $0 \le h_{\min} < h_{\max} \le 1$. The particle filter approximates the posterior distribution of the radar's threshold \( h_{\text{rad}} \) given the observations \( \tilde{g}_{1:t} \). The online threshold estimate is the posterior mean
\(
\hat h_{\text{rad},t} \;=\; \sum_{i=1}^N w_t^i\, h^i,
\)
while the stopping statistic \(\lambda_t\) is computed exactly as in \eqref{eq:lambda_expression}. As successive action observations are incorporated, the particle weights are updated using the action-observation likelihood. Consequently, particles associated with threshold hypotheses \(h^{(i)}\) that are closer to the true radar threshold tend to assign higher likelihood to the observed action sequence and therefore receive larger posterior weights, while less plausible hypotheses are progressively down-weighted. As a result, the posterior mean estimate \(\hat h_{\rm rad,t}\) adapts online as additional evidence becomes available. Although no theoretical convergence guarantee is established in this work, the numerical results demonstrate that the resulting threshold estimates are sufficiently accurate for reliable inverse detection.

We now present the algorithm implementing IQCD at the target with an unknown radar threshold. We replace the explicit indicators in \eqref{eq:prob_approximation}–\eqref{eq:prob_tau_approximation} with per-particle running Booleans \texttt{B}\(_t^i\) and \texttt{B}\(_\tau^i\) that indicate whether 
\(\pi_{t'}^i < h_{\text{rad}}^i\) holds for all \(t' \le t\) and for all \(t' \le \tau\), respectively 
Here “\(\land\)” denotes logical AND (conjunction), applied per particle to these Boolean flags.

\subsection{Uncertainty in the Radar and Action Observation Models at the Target: MM-IQCD}\label{subsection:MMIQCD}

The preceding development assumes that the target knows the radar's pre-change and post-change observation distributions \(f_0\) and \(f_1\), together with the target's action-observation distributions \(p_0\) and \(p_1\). In practice, the radar's observation model and the action-observation model may be unknown. 
The resulting uncertainty in \(f_0\), \(f_1\), \(p_0\) and \(p_1\) can be handled within the same multiple-model (MM) particle filtering framework described next.

Let \( \mathcal{M} = \left\{ \mathcal{M}^{(1)}, \mathcal{M}^{(2)}, \ldots, \mathcal{M}^{(M)} \right\} \) denote a finite collection of candidate models. Each candidate model is represented by
\(\mathcal M^{(j)}
=
\big(f_0^{(j)},f_1^{(j)},p_0^{(j)}, p_1^{(j)}\big).
\)

The target runs one particle filter for each candidate model
\(\mathcal M^{(j)}\), using the particle propagation, weight update,
normalization, and resampling steps described in
Section~\ref{section:PF}. The only modification is that the candidate
model is used in the corresponding stage of the filter.

If the radar observation model is uncertain at the target, the particle
propagation step employs the candidate likelihood ratio
\(
L_t^{(i,j)}
=
\frac{f_1^{(j)}(Y_t^{(i,j)})}
     {f_0^{(j)}(Y_t^{(i,j)})},
\)
whereas the remaining particle filter equations remain unchanged.

Similarly, if the action-observation model is uncertain at the target, the particle
weights are updated using the corresponding candidate likelihood
\(p(\tilde g_t|\pi_t^{(i,j)},\mathcal M^{(j)})\).
After normalization, each candidate filter computes its own inverse
detection statistic, denoted by \(\lambda_t^{(j)}\), using the same
procedure as in Section~\ref{section:PF}.

Thus, MM-IQCD preserves the original inverse particle filtering framework, with model uncertainty handled through parallel candidate filters whose outputs are combined according to their posterior model probabilities.

Let \(\rho_j\) denote the prior probability assigned to candidate model \(\mathcal{M}^{(j)}\). The posterior model probabilities are obtained from the Bayes theorem as
\begin{equation}
q_t^{(j)}
=
P\left(
\mathcal{M}=\mathcal{M}^{(j)}
\mid
\mathcal{I}_t
\right)
=
\frac{
\rho_j \mathcal{L}_t^{(j)}
}{
\sum_{m=1}^{M}
\rho_m \mathcal{L}_t^{(m)}
},
\end{equation}
where
$\mathcal{L}_t^{(j)}=P\big(\mathcal{I}_t\mid\mathcal{M}^{(j)}\big)=P\big(\tilde g_{1:t}\mid\mathcal{M}^{(j)}\big)$, where the target's state trajectory \(X_{1:t}\) is known and hence
treated as fixed, and \(\mathcal{L}_t^{(j)}\) denotes the marginal
likelihood of the observations under candidate model
\(\mathcal{M}^{(j)}\). A uniform prior, \(\rho_j=1/M\), is used when no candidate model is preferred a priori.

To evaluate the marginal likelihood, consider the action observations
\(\tilde g_{1:t}=\{\tilde g_1,\ldots,\tilde g_t\}\). For candidate
\(\mathcal{M}^{(j)}\), the one-step predictive likelihood of the current
action observation is $P\big( \tilde g_t \mid \tilde g_{1:t-1}, \mathcal{M}^{(j)} \big).$
The particle filter approximates this quantity using the weighted
particles as
\begin{equation}
c_t^{(j)}
=
\sum_i
w_{t-1}^{(i,j)}
p\left(
\tilde g_t
\mid
\pi_t^{(i,j)},
\mathcal M^{(j)}
\right).
\end{equation}
Thus, \(c_t^{(j)}\) represents the particle-filter estimate of the
one-step predictive likelihood under candidate model \(j\).

The marginal likelihood of the complete action-observation sequence
is obtained by the chain rule:
\[
P\left(
\tilde g_{1:t}
\mid
\mathcal{M}^{(j)}
\right)
=
\prod_{s=1}^{t}
P\left(
\tilde g_s
\mid
\tilde g_{1:s-1},
\mathcal{M}^{(j)}
\right).
\]
Therefore, its particle-filter approximation is
\begin{equation}
\hat{\mathcal L}_t^{(j)}
=
\prod_{s=1}^{t}
c_s^{(j)}.
\end{equation}
This product follows from the sequential factorization of the joint
likelihood and does not require the action observations to be independent.

The posterior model probabilities are then approximated using the
particle-filter estimate \(\hat{\mathcal L}_t^{(j)}\) as
$q_t^{(j)}
\approx
\frac{
\rho_j \hat{\mathcal L}_t^{(j)}
}{
\sum_{m=1}^{M}
\rho_m \hat{\mathcal L}_t^{(m)}
}.$

Rather than selecting a single candidate model, the target marginalizes over the model uncertainty using the posterior model probabilities and forms the model-averaged inverse statistic,
\begin{equation}
\lambda_t^{\mathrm{MM}}
=
\sum_{j=1}^{M}
q_t^{(j)}
\lambda_t^{(j)}.
\end{equation}
This constitutes the proposed multiple-model IQCD (MM-IQCD).

The corresponding stopping rule is
\begin{equation}
T_{\mathrm{target}}
=
\inf
\left\{
t\ge 1:
\lambda_t^{\mathrm{MM}}
\ge h_{\text{tar}}
\right\}.
\end{equation}
When the candidate set contains a single model ($M=1$), MM-IQCD reduces to the original IQCD detector developed in the previous section.
The posterior model probabilities can also be used to estimate uncertain model parameters. For example, if the candidates differ through a scalar parameter \(\theta_j\), such as the mean of \(f_1\) or the mean of \(p_1\), then
\( \hat{\theta}_t^{\mathrm{MM}}
=
\sum_{j=1}^{M}
q_t^{(j)}
\theta_j. \)

\section{Numerical Results}\label{section:numerical_results}

\subsection{Simulation Setup}\label{subsec:setup}
We evaluate the proposed inverse QCD framework from the target's perspective under the following model.
The radar observes 
$Y_t \sim \mathcal{N}(0,1)$ pre–change and $Y_t \sim \mathcal{N}(1.1,1)$ post–change, forms a posterior statistic $\pi_t$, and stops when $\pi_t \ge h_{\text{rad}}$.
The radar's geometric prior is $p=0.035$, while the target adopts a nominal $\hat p=0.1$ for its Bayesian update.
At the target, the radar's actions are observed noisily as
\begin{align}\label{eq:sim_action_model}
\tilde g_t \sim 
\begin{cases}
p_0 \equiv \mathcal{N}(0,1), & \pi_t < h_{\text{rad}},\\
p_1 \equiv \mathcal{N}(1.1,1), & \pi_t \ge h_{\text{rad}},
\end{cases}
\end{align}
which induces the likelihood pair $p_0(\cdot)$ and $p_1(\cdot)$ in \eqref{eq:observation_model}.
Unless stated otherwise, the particle filter uses $N=700$ particles and we report averages over $m=2000$ Monte Carlo runs.

\subsection{Baseline Comparison with Known Models}\label{subsec:methods}
We first consider the setting in which the statistical models are known at the target and compare the proposed IQCD detector with two benchmark methods, BLIND-PF and SU-CUSUM. For IQCD, we consider both known and unknown radar detection thresholds.
\begin{enumerate}
\item \textbf{IQCD:} The target runs the particle filter from Sec.~\ref{section:target's IQCD} and~\ref{section:PF} using the action model \eqref{eq:sim_action_model}. Weights are updated as in \eqref{eq:weight_update}, Boolean indicators \(\texttt{B}_{\tau}^i,\texttt{B}_t^i\) are maintained as in Algorithm~\ref{alg:iqcd_online_joint}, and the stopping statistic is $\lambda_t$ from \eqref{eq:lambda_expression}. We consider both \emph{known} and \emph{unknown} $h_{\text{rad}}$.
\item \textbf{BLIND-PF:} In this technique we have an identical state propagation as in the case of IQCD, but the filter ignores the radar's action. Specifically, we set
$w_t^{i}=\tfrac{1}{N}$ for all $t,i$ and use the estimate
\[
\hat{\pi}_t=\sum_{i=1}^N w_t^{i}\pi_t^{i}=\tfrac{1}{N}\sum_{i=1}^N \pi_t^{i}.
\]
The stopping rule used is a threshold policy where: 
\[
\hat{T}_B=\inf\{t\ge1:\ \hat\pi_t\ge h_{\text{tar}}\},
\]
with $h_{\text{tar}}$ calibrated to the desired PFA.
Note that BLIND-PF does not use $\tilde g_t$ nor requires $p_0,p_1$ (action model), but it uses the target’s known $X_t$ and $\tau$ in the propagation of $\pi_t^{i}$ (via $f_0,f_1$ and $\hat p$). 
\item \textbf{Self-Unaware CUSUM (SU-CUSUM):} A standard CUSUM test \cite{lorden1971procedures} runs directly on $\{\tilde g_t\}$, unaware of the target's  state $X_t$ and the change instant $\tau$. 
With $S_0=0$, we define the detection statistic as:
\[
S_t \;=\; \max\{0,\; S_{t-1} + \log l_t\}, 
\quad
l_t \;=\; \frac{p_1(\tilde g_t)}{p_0(\tilde g_t)}.
\]

\noindent The CUSUM test then becomes: 
\[
\hat{T}_C = \inf\{t\ge1: \ S_t\ge h_{\text{tar}}\}.
\]
\end{enumerate}
We calibrate \(h_{\mathrm{tar}}\) by evaluating a fine uniform grid of candidate thresholds over the interval \([0,1]\), which is the support of the inverse detection statistic \(\lambda_t\). For each candidate threshold, the conditional false alarm probability is estimated empirically from Monte Carlo simulations. The calibrated threshold is chosen as the smallest value satisfying
\[
\widehat{\mathrm{PFA}}(h_{\mathrm{tar}})
\le
\hat{\alpha},
\]
where \(\widehat{\mathrm{PFA}}(h_{\mathrm{tar}})\) denotes the empirical estimate of
\(\mathbb{P}(\hat{T}<T\,|\,T\ge\tau)\).
The corresponding mean detection delay is then evaluated using the calibrated threshold. Unless stated otherwise, the search is performed over 5001 uniformly spaced threshold values.

\begin{figure}[!t]
    \begin{centering}
    \begin{center}
    \includegraphics[height=6.5cm, width=1\linewidth]{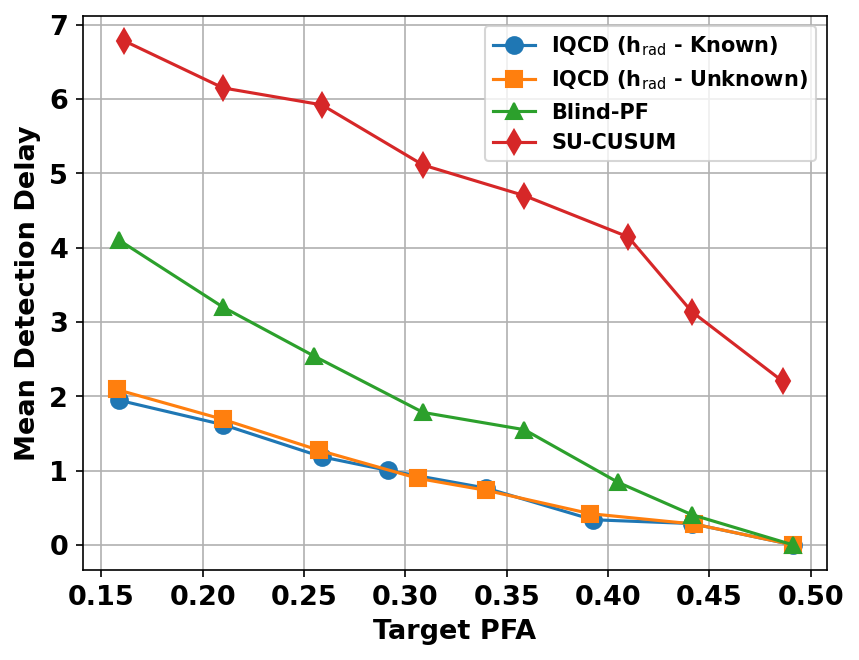}
    \end{center}
    \end{centering}
    \vspace{-5mm}
    \caption{Detection delay vs PFA at the target.}
    \label{fig:target_det_delay}
    \vspace{-2mm}
\end{figure}

Figure~\ref{fig:target_det_delay} plots the mean detection delay versus the probability of false alarm. The key observations are as follows:
\begin{itemize}
\item \textbf{IQCD (known vs. unknown $h_{\text{rad}}$):} The results indicate that online threshold inference does not considerably degrade detection performance. The near overlap between IQCD (known vs.\ unknown $h_{\mathrm{rad}}$) occurs because the SMC weights quickly concentrate around the true threshold and the ratio defining $\lambda_t$ is robust to small threshold errors.
\item Ignoring actions $\tilde g_t$ in BLIND-PF leads to noticeably larger delays, especially at low PFAs. This confirms that using the action likelihood (via \eqref{eq:weight_update}) is necessary.
\item  Operating only on $\tilde g_t$ without modeling the radar’s posterior as in SU-CUSUM yields the largest delay.
\end{itemize}
This ordering follows from how much information each method exploits. Overall, IQCD provides the best delay–false-alarm trade-off among the three detectors by exploiting both the target state information and the observed radar actions.
\begin{figure}[!h]
    \begin{centering}
    \begin{center}
    \includegraphics[height=6.5cm, width=1\linewidth]{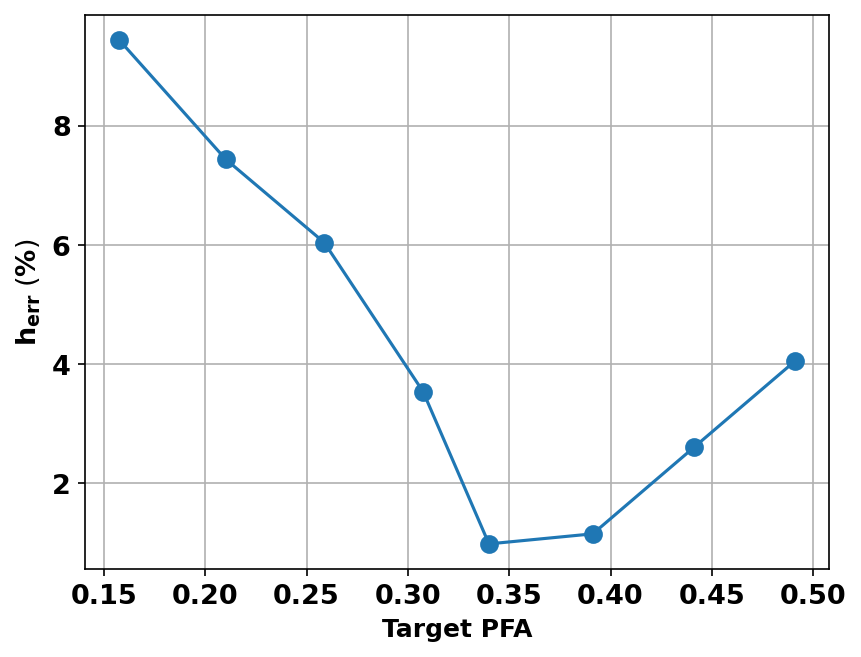}
    \end{center}
    \end{centering}
    \vspace{-5mm}
    \caption{Absolute estimation error between true and estimated radar thresholds}
    \label{fig:est_err}
    \vspace{-2mm}
\end{figure}
\subsection{Threshold-Estimation Accuracy}\label{subsec:threshold_accuracy}
Figure~\ref{fig:est_err} shows the relative estimation error,
\[
h_{\mathrm{err}}( \%) \;=\; \frac{\lvert \hat{h}_{\mathrm{rad}} - h_{\mathrm{rad}} \rvert}{h_{\mathrm{rad}}}\times 100,
\]
plotted against the target PFA in the unknown–threshold IQCD  setting.
The error remains small (typically $\lesssim 8\%$), which is consistent with the near-coincident delay curves for known/ unknown $h_{\mathrm{rad}}$ in Fig.~\ref{fig:target_det_delay}.
Although the estimated threshold is not exact, the estimation error remains sufficiently small that the inverse detection performance is nearly unchanged. This indicates that the proposed IQCD framework is relatively insensitive to moderate threshold estimation errors.

Thus, the particle filter-based target detection method shows consistent delay performance in both known and unknown radar threshold cases, despite slight differences in the estimated and true threshold values. This robustness can be attributed to the following key factors:
\begin{itemize}
    \item \textbf{Particle filter weight normalization:} Small errors in $\hat{h}_{\text{rad}}$ only slightly perturb the weights across particles. Since weights are normalized at each step, the overall distribution, and hence the timing of threshold crossing, remain relatively unaffected.
    \item \textbf{History dependent \( \lambda_t \) calculation:} The detection statistic \(\lambda_t\) depends on history of observations rather than a single threshold value. Minor threshold inaccuracies rarely alter the time at which \(\lambda_t\) exceeds \(h_{\text{tar}}\).
    \item \textbf{Robustness to threshold uncertainty:}
    The inverse detector depends on the posterior distribution of the radar state accumulated over time rather than on a single threshold estimate. Consequently, moderate inaccuracies in the estimated radar threshold have only a limited effect on the evolution of the detection statistic.
\end{itemize}
In summary, the similarity in delays between the known and unknown radar threshold cases highlights the resilience of the particle filter approach. By leveraging a large number of particles and normalization in weight calculation, the filter maintains consistent detection performance even under threshold uncertainties.

\subsection{Performance under Model Uncertainty}\label{subsec:sensitivity-analysis}
\begin{figure}
    \begin{centering}
    \begin{center}
    \includegraphics[height=6.5cm, width=1\linewidth]{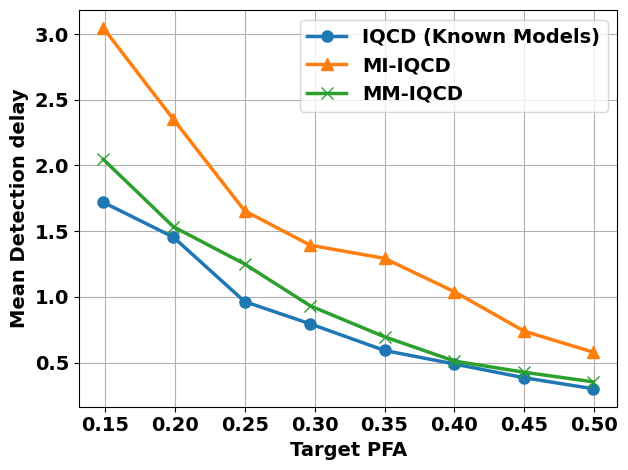}
    \end{center}
    \end{centering}
    \vspace{-5mm}
    \caption{Detection delay vs PFA for unknown radar observation and action models at the target.}
    \label{fig:target_det_delay_unknown_params}
    \vspace{-2mm}
\end{figure}

To evaluate robustness against uncertainty in the radar and action-observation models at the target, Figure~\ref{fig:target_det_delay_unknown_params} compares the proposed MM-IQCD with the original IQCD (with known $h_{\text{rad}}$) and a mismatched-model IQCD detector (MI-IQCD), in which the target incorrectly assumes a single fixed post-change model instead of performing model averaging. 

The true radar observation and action-observation models are as specified in Section~\ref{subsec:setup}. In MM-IQCD, however, the target assumes that \(f_0\), \(f_1\), \(p_0\) and \(p_1\) are unknown.  To handle this uncertainty, a finite candidate set is constructed using discretized grids. Specifically, the candidate values for the Gaussian means of $f_1$ and $p_1$ are drawn from $\{0.5, 0.75, 1.0, 1.25, 1.5, 2.0\}$, while those for $f_0$ and $p_0$ are drawn from $\{-0.50, -0.25, 0.10, 0.30, 1.0, 1.5\}$. The variances are assumed to be known and are fixed at $\sigma^2 = 1$. Assuming independent parameter uncertainties, taking all combinations of these parameter grids yields a model space of $M = 6^3 = 216$ candidate models. The candidate-specific inverse statistics are then combined using the posterior model probabilities described in Section~\ref{subsection:MMIQCD}.
\begin{figure}[!t]
    \begin{centering}
    \begin{center}
    \includegraphics[height=6.5cm, width=1\linewidth]{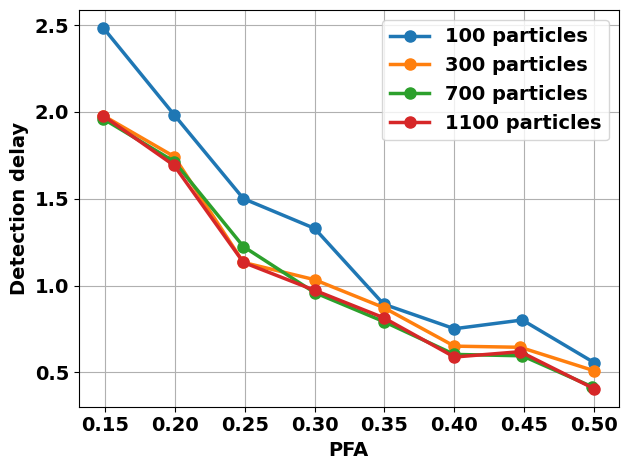}
    \end{center}
    \end{centering}
    \vspace{-5mm}
    \caption{Detection delay vs PFA at the target for various number of particles under MI-IQCD.}
    \label{fig:particle_sensitivity_unknown}
    \vspace{-2mm}
\end{figure}
For comparison, we also consider a mismatched-model IQCD detector (MI-IQCD), in which the target assumes a single fixed post-change configuration where $f_0$, $f_1$, $p_0$ and $p_1$ are modeled as Gaussian distributions with mismatched means of $1.9$, $3.5$, $1.6$ and $3.2$, respectively, whereas the true distributions have means of $0$, $1.1$, $0$ and $1.1$.

The results show that MM-IQCD achieves a probability-of-false-alarm versus delay trade-off that remains very close to the fully informed IQCD detector despite the uncertainty in both the radar post-change model and the action-observation model. This demonstrates that the proposed multiple-model framework effectively compensates for moderate model uncertainty by combining information from several candidate models rather than relying on a single nominal model. In contrast, MI-IQCD exhibits a noticeable increase in detection delay, particularly in the low-PFA regime, illustrating the performance degradation caused by persistent model mismatch.

Figure~\ref{fig:particle_sensitivity_unknown} shows the sensitivity of the detector to the number of particles in the MI-IQCD setting. The PFA-delay curves for $N \in \{100,300,700,1100\}$ are largely consistent across the considered PFA range, with the $N=100$ case showing a somewhat larger deviation at low PFA. Increasing the number of particles beyond $N=300$ results in only minor changes in the observed performance. These results indicate that the detector is reasonably stable with respect to the number of particles.

\subsection{Sensitivity to the Nominal Prior Parameter}
We next examine robustness of IQCD to misspecification of the geometric prior parameter used by the target. The radar employs the true parameter \(p=0.035\), whereas the target propagates its particles using a nominal value
\(
\hat p
\in
\{0.02,\,0.035,\,0.05,\,0.075,\,0.10,\,0.15\}.
\)
For each value of $\hat p$, $h_{\text{tar}}$ is calibrated by empirically evaluating the false declaration probability in \eqref{eq:target_optimization_problem} over a grid of thresholds and selecting the smallest threshold satisfying the prescribed constraint. Thus, the comparison represents the effect of prior-parameter mismatch on the detection delay for the same false-alarm rate.

\begin{figure}[!t]
    \begin{centering}
    \begin{center}
    \includegraphics[height=6.5cm, width=1\linewidth]{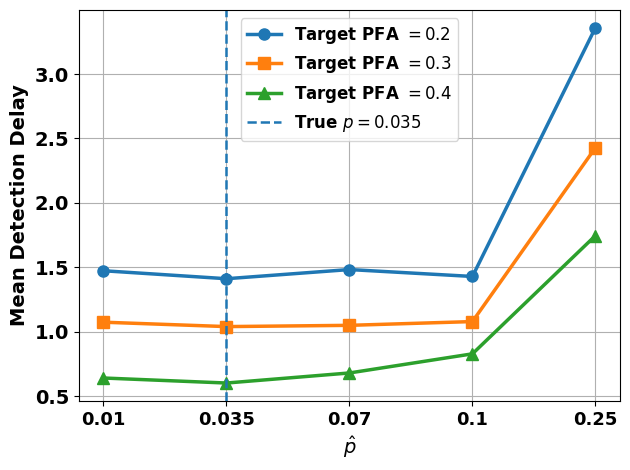}
    \end{center}
    \end{centering}
    \vspace{-5mm}
    \caption{Detection delay vs $\hat p$ at the target for various PFA values.}
    \label{fig:p_mismatch}
    \vspace{-2mm}
\end{figure}

Figure~\ref{fig:p_mismatch} shows the resulting mean detection delay vs $\hat p$ for varying PFA values. The results indicate that IQCD remains effective under moderate misspecification of the prior parameter, although larger deviations from the true value lead to an increased detection delay. Moderate mismatch in the geometric prior parameter has only a minor impact on detection performance because the particle weights are continually corrected using the observed radar actions, thereby compensating for inaccuracies introduced during the prediction step.
\subsection{Statistic Trajectories}\label{subsec:trajectories}
\begin{figure}[!h]
    \begin{centering}
    \begin{center}
    \includegraphics[height=6.5cm, width=1\linewidth]{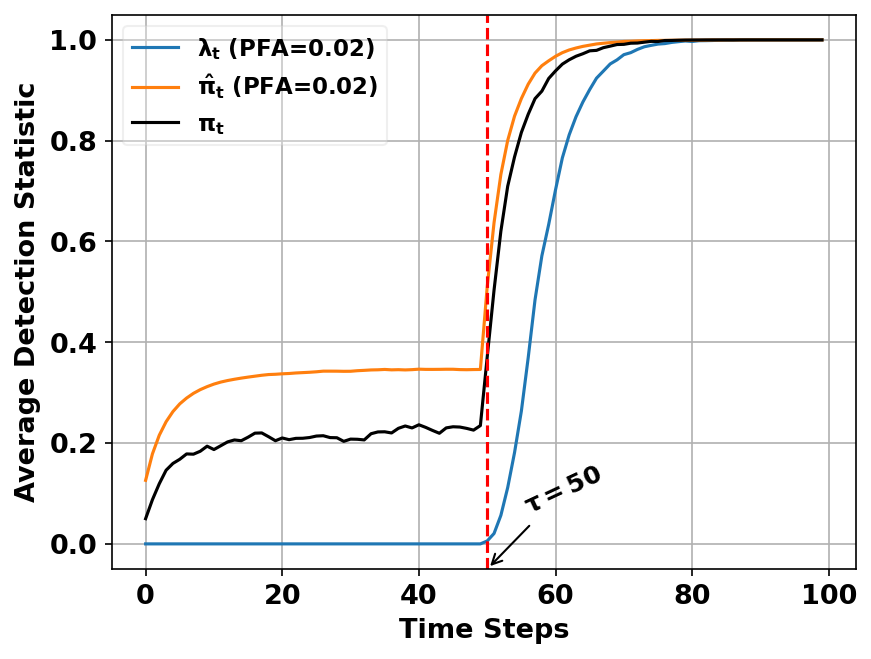}
    \end{center}
    \end{centering}
    \vspace{-5mm}
    \caption{Average trajectories of detection statistics over all sample paths (representative case with target PFA $=0.02$).}
    \label{fig:detec_stat}
    \vspace{-2mm}
\end{figure}

Figure~\ref{fig:detec_stat} illustrates temporal variations over $m=2000$ runs at target PFA $=0.02$ for $\lambda_t$ (IQCD), $\hat{\pi}_t$ (BLIND\textendash PF), and the radar’s true $\pi_t$. The results shown correspond to a representative setting with target PFA $=0.02$. The vertical dashed line marks the change time $\tau=50$. All three curves stay below $0.4$ before $\tau$ and rise smoothly toward $1$ by $t\approx60$. In our Gaussian setup, BLIND–PF’s $\hat{\pi}_t$ aligns closely with the average radar statistic $\pi_t$ across runs. To be precise,  $\hat{\pi}_t$ tracks $\pi_t$ well in expectation (over paths). On individual paths or under model mismatch, the agreement can be weaker, and ignoring $\tilde g_t$ typically yields longer detection delays. Note that the trajectories of $\lambda_t$, $\hat{\pi}_t$, and $\pi_t$ are independent of the target PFA; the PFA only influences the choice of the detection threshold applied to these statistics.

Across all sample paths, IQCD outperforms SU-CUSUM and clearly dominates BLIND-PF. Moreover, unknown $h_{\text{rad}}$ IQCD matches the known-threshold performance while keeping threshold inference fully online. These results demonstrate that inverse QCD with particle filtering achieves good detection performance despite the noisy action-observation model and uncertainty in the radar detection threshold, enabling accurate and timely estimation of the radar's stopping time.

\section{Conclusion}\label{section:conclusion}

In this paper, we presented an inverse Bayesian quickest change detection (QCD) framework from the target’s perspective, wherein the target seeks to infer the radar’s stopping time using noisy observations of its actions. We formulated the inverse QCD problem to account for both known and unknown threshold scenarios, and proposed a particle filtering approach for real-time estimation of the radar’s posterior and stopping time. We further extended the framework to a multiple-model IQCD (MM-IQCD) formulation that accounts for uncertainty in the radar's observation model and the action-observation models at the target. Numerical results demonstrated that the proposed methods accurately track the radar’s stopping time and remain robust under moderate model uncertainty and parameter mismatch, thereby enabling the target to adopt suitable countermeasures against the radar's actions. Our work lays the foundation for the development of a systematic theory of inverse QCD in the broader context of inverse cognition.

Important directions for future research include establishing asymptotic performance guarantees for the proposed inverse QCD framework, analogous to those available for classical Bayesian QCD, and developing computationally efficient alternatives to MM-IQCD. Since MM-IQCD requires running one particle filter for each candidate model, its computational complexity increases approximately linearly with the number of candidate models. Generalized likelihood ratio (GLR)-based approaches may offer attractive low-complexity alternatives while retaining robustness to model uncertainty.

\bibliographystyle{IEEEtran}
\bibliography{bibreport}

\end{document}